\documentclass[12pt]{iopart}

\usepackage{graphicx}

\begin{document}

\title[Effects of Black Hole Singularities on Scalar Fields]
{Effects of Schwarzschild's Black Hole Singularities on Complex Scalar Fields}

\author{Z.E. Musielak, J.L. Fry and G.W. Kanan} 

\address{Department of Physics, The University of Texas 
at Arlington, Arlington, TX 76019, USA}

\ead{zmusielak@uta.edu (Corresponding author)}

\begin{abstract}
Complex scalar fields described by a novel Klein-Gordon 
equation derived from gauge and group theories are 
considered at the Schwarzschild's black hole singularities.  
It is shown that the field is well-behaved in the vicinity 
of these singularities and that its value reaches zero at 
both singularities.  The obtained results also demonstrate 
that the field forms a scalar hair that exists outside of 
the event horizon, and that the interior field is tachyonic 
and undergoes a tachyonic condensation to reach its true 
vacuum at the central singularity.  The described field's 
behavior is very different from that predicted by the 
Klein-Gordon equation minimally coupled to gravity.  
Physical implications of these results for the interior 
structure of black holes are discussed. 
\end{abstract}

%
%

\section{Introduction}

The fundamental relativistic equation that describes evolution 
of complex or real scalar fields on the Minkowski metric is 
the Klein-Gordon (KG) equation [1,2].  A detailed description 
of the KG equation and its applications can be found in any 
quantum field theory (QFT) textbook (e.g., [3,4]).  The 
KG equation is typically obtained by using the relativistic 
energy-momentum relationship, which is based upon 
classical equations for point masses (e.g., [5]), and by
 replacing its four-momentum by the four-momentum 
operator (e.g., [6]).  Symmetry analysis of the KG equation
 is presented and discussed in [7].  It was also demonstrated 
that the irreducible representations (irreps) of the Poincar\'e 
group [8,9] can be used to identify and classify the fundamental 
equations of QFT, which include the KG equation [10].

Another method to derive the KG equation is to use the eigenvalue 
equation for the energy-momentum operator, whose time and 
space translation operators are the generators of $T (3+1)$ that 
is an invariant subgroup of the Poincar\'e group [9].   The action 
of these operators on a complex scalar field gives the corresponding 
eigenvalues [11] that were used to derive the KG equation [12] as 
well as its higher-derivative forms [13] in flat spacetime.  However, 
the complexity of the problem dramatically increases when spacetime
 is curved, and its metric tensor is described by Einstein's General 
Relativity (GR) (e.g., [14-18]).  The main difficulty is to identify the 
form of the KG equation that would correctly represent the evolution 
of a complex scalar field in curved spacetime of a given metric.

Different forms of the KG equations in curved spacetime have been 
proposed (e.g., [14-23]).  Typically, the KG equation in flat spacetime 
is modified by replacing its partial derivatives by covariant derivatives, 
keeping the mass term constant, which is called the minimal coupling 
to gravity.  In some studies, a curvature term that contains the Ricci 
scalar multiplied by a numerical factor was also used to modify the 
KG equation.  This curvature term was formally introduced by 
imposing conformal invariance on the resulting KG equation 
[14,16-18], and the term describes a hypothesized coupling 
between the scalar and gravitional fields.  The KG equation 
with this term was used to formulate a quantum field theory 
(QFT) in globally hyperbolic spacetimes [14-18].  In the 
formulated QFT, no reference was made to the plane-wave 
basis, but instead the so-called 'operator-valued distributions' 
were used with no preferred choice of Hilbert space, which 
resulted in no uniquely defined vacuum in the theory (e.g., 
[15-18]).  For given metrics, such as the Schwarzschild or 
Kerr or Robertson-Walker metric, typically the massless KG 
equation was used (e.g., [18,24,25]). 

In recent work [26], a gauge theory of a complex scalar field 
in curved spacetime was developed and its main results were 
used to obtain a {\it metric-dependent Klein-Gordon (md-KG) 
equation}.  The derived md-KG equation significantly differs 
from the KG equations described above.  In this paper, we 
begin with a brief summary of this gauge theory and then 
derive the same md-KG equation by using group theory.  
The md-KG equation in curved spacetime describes temporal 
and spatial evolution of the scalar field on pseudo-Riemannian 
manifolds using only properties of the metric tensor.  It contains 
no adjustable parameters and relies on mathematically rigorous 
properties of tensor fields on arbitrary metric manifolds. 

Scalar fields on manifolds were previously investigated by several 
authors (e.g., [27-29]), but the context of those studies is not 
directly related to the results presented in this paper.  However,
investigations of scalar fields around black holes that resulted in
the black holes with the so-called scalar hair (e.g., [30], and 
references therein) are relevant to this paper.  Specifically, the 
most relevant previous work is on growth of massive [31] or 
massless [32] scalar hair around a Schwarzschild black hole,
but also studies of hairy charged black holes (e.g., [33], and 
references therein).  It must be noted that in that previous 
work on massive scalar fields, the minimally coupled massive 
KG equation was used, and that this equation is very different 
from the md-KG equation, which describes the behavior of 
complex scalar field in the Schwarzschild spacetime.  

In this paper, we investigate the effects of Schwarzschild's black 
hole singularities on physical properties of complex scalar fields 
near and at the black hole's event horizon and central singularities 
(e.g., [14,34-37]) by using the md-KG equation.  The obtained 
results demonstrate that the field's behavior is very different 
outside and inside the black hole, and that the field forms 
a scalar hair that exists near the event horizon.  Moreover, 
it is shown that the interior field is tachyonic and as a result 
the field rolls down from its unstable maximum at the event 
horizon to its true minimum located at the central singularity.

This paper is organized as follows.  Section 2 presents our derivation 
of the metric-dependent Klein-Gordon equation in curved spacetime;  
solutions to the metric-dependent Klein-Gordon equation for the 
spherical and Lemaitre coordinates in the Schwarzschild metric at 
its singularities are given and discussed in Section 3; physical 
implications of the obtained results are discussed in Section 4; 
and our conclusions are given in Section 5.

\section{Metric-dependent Klein-Gordon equation in curved spacetime}

\subsection{Basic setting and observers in curved spacetime}

Curved spacetime of GR is described by a 4D, smooth, pseudo-Riemannian 
manifold ${\cal M}$ endowed with the metric $ds^2 = g_{\mu \nu} (x)\ 
dx^{\mu} dx^{\nu}$, where $g_{\mu \nu}$ is the metric tensor, $[x] \equiv 
[x^{\mu}]$, and $\mu$ and $\nu$ are $0$, $1$, $2$ and $3$, and the usual 
summation conventions are employed.  GR replaces the flat spacetime of 
Special Relativity (SR) by ${\cal M}$ and requires that the Poincar\'e group, 
which is the group of transformations that leave dynamical equations invariant 
in SR, be replaced by a general group of coordinate transformations known as 
the diffeomorphism group and denoted as Diff$({\cal M})$ (e.g., [38,39]).  
This group is defined on the entire manifold, and it is a spatio-temporal 
group that carries more limited information than the Poincar\'e group. 
    
Let us introduce two observers whose coordinate systems may be related 
by linear and continuous transformations, one with coordinates $x$ and 
another with $x^{\prime}$.  Using this definition of the observers, the 
principle of general covariance, which was used by Einstein as a basis for 
developing GR, can be introduced [14].  A modern view is that general 
covariance is a tool but not a principle.  Specifically, Weinberg [34] and 
others [40,41] claim that 'the principle of general covariance is not an 
invariance principle'.  However, since the terms in the md-KG equation 
derived in this paper are scalars, we shall refer to this equation as 
invariant.    

\subsection{Gauge theory of quantum scalar fields in curved spacetime}

We introduce a complex scalar field $\Phi (x)$ on a smooth and 
curved pseudo-Riemanian manifold ${\cal M}$, with the metric 
tensor $g_{\mu \nu} (x)$.  We use the metric signature 
$(+, -, -, -)$, and impose the usual assumption that all 
observers, who are related by diffeomorphism, identify the 
same field and agree upon its physical nature and description.  

To develop a gauge theory, we follow [26] and specify a null 
Lagrangian density (NLD) for the complex field $\Phi (x)$ as 
${\cal {L}}_{n} = \nabla_{\mu} [ q^{\mu} (x) \Phi (x) 
\Phi^{*} (x) ]$, or 
\begin{equation}
{\cal {L}}_{n} = [ \nabla_{\mu} q^{\mu} (x) ]\ \Phi (x) 
\Phi^{*} (x) + q^{\mu} (x) \nabla_{\mu} [ \Phi (x) ) 
\Phi^{*} (x)]\ ,
\label{eq1}
\end{equation}
where $q^{\mu} (x)$ is assumed a real field to be determined, 
and $\Phi^{*} (x)$ is a complex conjugate of $\Phi (x)$.  Thus, 
${\cal {L}}_{n}$ is a real function.  Substitution of ${\cal {L}}_{n}$ 
into the Euler-Lagrange (E-L) equations for the variations in either 
$\delta \Phi (x) $ or $\delta \Phi^{*}$ or $\delta q^{\mu}$ (see 
below for its definition) makes these equations identically zero 
(e.g., [42-44]).  Since the NLD does not give any dynamical 
equation for $\Phi (x)$, there are no predetermined constraints 
on this scalar field.  

We introduce the following local gauge transformation [26]
\begin{equation}
\label{}
\Phi (x) \rightarrow \Phi (x) \exp \left( -i \int^x p_\mu (\tilde{x}) 
d\tilde{x}^\mu \right),
\label{eq2}
\end{equation}
where the integrand $p  = p_\mu (x) dx_\mu$ is a real one-form on 
the manifold, with $p_\mu$ and $dx^\mu$ being its components and 
basis of $p$, respectively.  For uniqueness of this transformation the 
integral must be path-independent.  This requires that the one-form 
be exact, which in turn requires that $p = d \chi(x)$, where $d$ is 
the exterior derivative and $\chi(x)$ is a scalar function.  Note that 
$d(d\chi) = d^2 \chi = 0$.  

The requirement that $p$ be exact does not determine the one-form 
$p$ nor its components, $p_\mu (x)$.  Some condition must be 
imposed on $p_\mu (x)$ to allow forits determination.  Consistent 
with the intent of this paper, the solenoidal condition $\nabla^\mu 
p_\mu (x) = 0$ is imposed, with $\nabla^\mu = g^{\mu \nu} (x) 
\nabla_\nu$, which for a given metric may be solved to obtain 
$p_\mu (x)$ [26].  The solenoidal condition will be true if $p_\mu 
(x)$ is exact, which means that $p_\mu = \nabla^\mu \chi(x)$ for 
some scalar function $\chi(x)$.  Since $\nabla_\mu \nabla^\mu 
\chi(x) = 0$, we can solve for $\chi(x)$ and, hence, determine 
$p_\mu(x)$.  

Now, the field $q^\mu (x)$ that was used to define the NLD has 
additional degrees of freedom given by $q^\mu (x) \rightarrow 
q^\mu (x) + \nabla^\mu \chi(x)$, where the second term is the
 variation of the field $q^\mu (x)$, i.e., $\delta q^\mu (x) = 
\nabla^\mu \chi(x) = p^\mu (x)$.  Operating on $\delta 
q^\mu$ with $\nabla_\mu$ gives $\nabla_\mu (\delta q^\mu) 
= \nabla_\mu \nabla^\mu \chi = 0$, and since $\nabla_\mu 
(\delta q^\mu) = \delta (\nabla_\mu q^\mu)$, we obtain 
$\nabla_\mu q^\mu = 0$ which shows that $q^\mu$ is 
also a solenoidal field.  Given a metric, the solenoidal 
property of $q^\mu$ allows for its solution.  Note that 
the condition $\nabla_\mu (\delta q^\mu) = \delta 
(\nabla_\mu q^\mu)$ is satisfied because there are 
no changes in the metric caused by either $q^{\mu} (x)$
or $p_{\mu} (x)$ or $\Phi (x)$ [45]. 

Applying the gauge transformation given by Eq. (\ref{eq2}) 
to Eq. (\ref{eq1}), the transformed NLD, ${\cal {L}}_n^{tr}$,
is obtained.  Then, the difference between the transformed 
and original NLDs is calculated ${\cal {L}}_n^{tr} - 
{\cal {L}}_n = \Delta {\cal {L}}_n$ (for details, see [26]).
The theory is gauge invariant if $\Delta {\cal {L}}_n = 0$, 
which gives the following local eigenvalue equations
\begin{equation}
- i \nabla_{\mu} \Phi (x) = p_{\mu}(x) \Phi (x)\ ,
\label{eq3}
\end{equation}
and
\begin{equation}
i \nabla_{\mu} \Phi^{*} (x) = p_{\mu}(x) \Phi^{*} (x)\ . 
\label{eq4}
\end{equation}
Note that by using the metric tensor, both equations 
can be converted into their contravariant forms.  The 
derived eigenvalue equations ensure the conditions 
for gauge invariance of action for the NLD.   

The fact that one-form $p = p_{\mu}(x) dx^{\mu}$ 
is exact guarantees that the simultaneous solutions to 
the eigenvalue equations and the solenoidal condition 
for $p_{\mu} (x)$ exist for a given metric.  The local 
eigenvalue equations together with the solenoidal 
condition were used in [26] to obtain the md-KG 
equation.  The simultaneous solutions also naturally 
satisfy the md-KG equation obtained for the metric.  
In the following, we demonstrate that the same 
local eigenvalue equations can be obtained using 
group theory.

\subsection{Group theory derivation of eigenvalue equations}

In the Minkowski spacetime with time $t$ and global Cartesian 
coordinates $(x, y, z)$, we have $x_C = (t, x, y, z)$, and the
irreps of the Poincar\'e group [8,9] can be used to establish 
local labels for the field $\Phi (x_C)$.  For this field to transform 
as one of the unitary irreps of the invariant subgroup $T(3+1)$ 
of the Poincar\'e group, $\Phi (x_C)$ must obey the eigenvalue 
equation $- i \partial_{\mu} \Phi (x_C) = k_{\mu} \Phi (x_C)$, 
where the eigenvalue $k_{\mu}$ = const.  Using this eigenvalue 
equation, the KG equation in Cartesian coordinates was derived 
[11,12].   

However, for curved pseudo-Riemannian manifold ${\cal M}$, the 
Poincar\'e group represents only local spacetime symmetries [9].
Thus, there are no global eigenfunctions of the space and time 
translation operators similar to those given above for flat spacetime.
Instead the scalar field $\Phi (x)$ has different gradients at different 
points on the curved manifold.  As a result, the global eigenvalue 
equations found in the Minkowski spacetime must be generalized 
to account for the curvature of ${\cal M}$.  We achieve this by 
replacing $k_{\mu}$ by a one-form $p_{\mu} (x)$.  With 
$k_{\mu} \rightarrow p_{\mu} (x)$, we convert the eigenvalue 
equation in flat spacetime into the following local eigenvalue 
equation
\begin{equation}
- i \nabla_{\mu} \Phi (x) = p_{\mu} (x) \Phi(x)\ ,
\label{eq5}
\end{equation}
\noindent
where $p_{\mu} (x)$ is a real vector field to be determined 
over the manifold.  We see that Eq. (\ref{eq5}) is of the 
same form as Eq. (\ref{eq3}) obtained from the gauge 
theory.  However, no other restrictions on $p_{\mu} (x)$ 
emerge from the group theory derivation.  Thus, since 
Eqs (\ref{eq3}) and (\ref{eq5}) are identical, with 
$p_{\mu} (x)$ being the local eigenvalue for both 
equations, we assume that $p_{\mu} (x)$ represents 
the components of the one-form $p$ and obeys the 
conditions on $p$ imposed by the gauge theory 
described above.  With this assumption, the gauge 
and group theories give the same results. 

\subsection{Derivation of metric-dependent Klein-Gordon equation}

To derive the md-KG equation in curved spacetime, we apply 
the operator $i \nabla^{\mu}$ to the local eigenvalue equation 
given by Eq. (\ref{eq3}), and use the solenoidal condition for 
$p_{\mu} (x)$.  Then, we obtain the following equation
\begin{equation}
\left [ \nabla^{\mu} \nabla_{\mu} + p^{\mu}(x) p_{\mu}
 (x) \right ] \Phi(x) = 0\ ,
\label{eq6}
\end{equation}
This md-KG equation is valid for any curved spacetime represented 
by a pseudo-Riemannian manifold on which the term $p^{\mu} 
(x) p_{\mu} (x)$ is defined.  For a given metric of curved spacetime, 
the explicit form of the term is determined by solving $\nabla^{\mu} 
p_{\mu} (x) = 0$, and with $p^{\mu} (x) = g^{\mu \nu} (x) 
p_{\nu} (x)$.    This metric-dependent term reduces to $p^{\mu} 
(x) p_{\mu} (x) = \Omega_o^2 = m_o^2 c^4 / \hbar^2$, where 
$m_o$ is the mass associated with the scalar field, in the limit 
($r \rightarrow \infty$) of asymptotically flat spacetime with the 
Minkowski metric.  This shows that in the limit of flat spacetime, 
the md-KG equation reduces to the KG equation well-established in 
QFT [3,4,6]. 

Comparison of the obtained md-KG equation to the previously 
used KG equations in curved spacetime (e.g., [14-23,30,31,33]) 
shows that the term $p^{\mu} (x) p_{\mu} (x)$ is significantly 
different from those considered before.  Typically, the terms 
previously used involved either the constant mass term, 
$\Omega_o^2$ = const [46-51], or the local value of the 
Ricci scalar multiplied by a numerical factor, which was 
introduced to the equation by postulating its conformal 
invariance [14,16].  In other studies (e.g., [18,24,25,32]), the 
massless KG equation was used.  In the approach presented 
here, the md-KG equation with its metric-dependent term 
$p^{\mu} (x) p_{\mu} (x)$ results directly from the 
eigenvalue equations, which are the conditions that the 
presented theory is gauge invariant and consistent with 
the local symmetries of pseudo-Riemannian manifolds 
represented by the Poincar\'e group.  Note that the 
derived md-KG equation is diffeomorphism invariant.

\section{Complex scalar field in the Schwarzschild metric}

\subsection{Metric-dependent Klein-Gordon equation in 
spherical coordinates}

To illustrate the basic properties of the md-KG equation (see Eq. \ref{eq6}), 
we consider a Schwarzschild black hole with its surroundings described 
by the Schwarzschild metric in spherical coordinates (e.g., [14,34-37]).  
In these coordinates $x = (t, r, \theta, \phi)$, and we introduce a quantum 
scalar field $\Phi (x)$ that is present in the entire spacetime, including the 
black hole's interior, and use the md-KG equation to investigate the behavior 
of the field in this environment.  To determine $p_{\mu} (x)$, the solenoidal 
condition for this field is solved by separation of variables in the Schwarzschild 
metric [26], and a solution is  
\begin{equation}
p_{\mu} (x) = \left ( K_t , \frac{K_r}{(r - R_S) r}, \frac{K_\theta}
{\sin \theta}, K_{\phi} \right )\ ,
\label{eq7}
\end{equation}
where $K_t$, $K_r$, $K_{\theta}$ and $K_{\phi}$ are the intergration 
constants to be determined by boundary conditions, and $R_S = 2 
G M / c^2$ is the Schwarzschild radius, with $M$ being black hole's 
mass.  The obtained solution guarantees that $p_{\mu} (x)$ is exact 
since separation of variables ensures that a scalar function $\chi (x)$ 
exists such that $p_{\mu} (x) = \partial_{\mu} \chi (x)$; see discussion 
below Eq. (\ref{eq2}).  Similarly, we may solve $\nabla_{\mu} q^{\mu} 
(x) = 0$ to obtain 
\begin{equation}
q^{\mu} (x) = \left ( K^t , \frac{K^r}{r^2}, 
\frac{K^\theta}{\sin \theta}, K^{\phi} \right )\ , 
\label{eq8}
\end{equation}
which shows the this vector field is non-zero and non-singular in 
the entire Schwarzschild spacetime, except at the central singularity.
As a result, the NLD that is dependent on $q^{\mu} (x)$ is also 
well-defined in the Schwarzschild metric, with $\theta \neq 0$, 
except its singularity at $r = 0$.  Note also that $g^{\mu \nu} 
(x) p_{\nu} (x) \neq q^{\mu} (x)$.

Using the metric tensor $g^{\mu \nu} (x)$ for the Schwarzschild 
metric, the field $p^{\mu} (x)$ can be calculated.  Then, in a rest 
frame associated with a stationary observer located at any $r > R_S$ 
in the Schwarzschild spacetime,  $K_r = K_{\theta} = K_{\phi} = 0$ 
but with the rest frame value of $K_t \neq 0$.  Then, the term 
$p^{\mu} (x) p_{\mu} (x)$ becomes
\begin{equation}
p^{\mu} (x) p_{\mu} (x) = \left ( 1 - \frac{R_S}{r} \right )^{-1} 
K_t^2\ ,
\label{eq9}
\end{equation}
where $p^{\mu} (x) = g^{\mu \nu} (x) p_{\nu} (x)$; note 
that $p^{\mu} (x) \neq q^{\mu} (x)$, with the latter given 
by Eq. (\ref{eq8}).   To determine the integration constant, 
$K_t$, we assume that in flat spacetime, the md-KG equation 
reduces to the KG equation well-known in QFT [3,4,6,7].  
Since in the limit $r \rightarrow \infty$, the Schwarzschild 
metric becomes the Minkowski metric, we choose $K_t^2 = 
\Omega_o^2 = m_o^2 c^4 / \hbar^2$, so that the md-KG 
equation becomes the KG equation in this limit (see Eq. 
\ref{eq6} and its discussion).   

Thus, the md-KG equation in the Schwarzschild metric can be 
written as
\begin{equation}
\left [ \nabla^{\mu} \nabla_{\mu} + \left ( 1 - \frac{R_S}{r} 
\right )^{-1} \Omega_o^2 \right ] \Phi (x) = 0\ ,
\label{eq10}
\end{equation}
and the factor $(1 - R_S/r)^{-1}$ guarantees that in the limit 
$r \rightarrow \infty$, the equation becomes the KG equation 
in Minkowski spacetime [3,4,6,7].  Since the md-KG equation 
is covariant with respect to the diffeomorphism group, its 
form remains the same for any coordinates used in this 
metric, including the Lemaitre, Eddington-Finkelstein 
and Kruskal-Szekeres coordinates [7,36].  However, the 
expression for the Laplace-Beltrami operator $\nabla^{\mu} 
\nabla_{\mu}$ has different forms in different coordinates
[34-36].

Let us point out that General Relativity (GR) can be formulated 
as a gauge theory, and that locally either the diffeomorphism 
group, or the Poincar\'e group, becomes a gauge group of 
spacetime symmetry [7].  This means that gauge freedom in
 GR refers to selection of a specific coordinate system, and 
that gauge transformations are coordinate transformations.  
As a result, metrics that look differently in different coordinate 
systems describe the same spacetime.  This is consistent with 
the above results and with the form of the md-KG equation, 
which is independent from different coordinates used in the 
same metric.  

Despite GR being a gauge theory, it significantly differs 
from QFT gauge theories.  The main difference is that 
the former uses one of the groups of spacetime symmetries 
[7].  However, the latter are based on internal symmetry 
unitary groups such as U(1), SU(2) and SU(3), or their 
combination [3,4].  Now, the gauge theory described in 
Section 2.2 is novel as its gauge transformation is based 
on components of one-form $p_{\mu} (x)$ that is local 
and determined by a given metric.  Moreover, this novel
theory requires specifying a null Lagrangian, which makes
the theory free of any predetermined physical conditions.  
Nevertheless, the resulting md-KG equation arises 
naturally from the theory, and it accounts for effects 
of spacetime curvature as described by the metric.

The md-KG equation given by Eq. (\ref{eq9}) has some interesting 
properties.  One of them is that the sign of the second term in 
the md-KG equation changes when $r$ becomes less than $R_s$. 
This implies that the interior and exterior solutions to the md-KG 
equation are different; actually, the interior solutions are 
'tachyonic' [51-57].  Moreover, when $r \rightarrow 0$, the 
second term in the md-KG equation approaches zero, which 
means that the scalar field $\Phi (x)$ becomes massless 
and well-behaved at the black hole's central singularity (see 
Section 5 for details and discussion).  This shows that the 
presence of the term $(1 - R_S / r)^{-1}$ in the md-KG 
equation separates the field $\Phi (x)$ into two differently 
behaving fields, one located inside the black hole where 
$r < R_S$, and the other outside where $r > R_S$.  
These effects were not present in previous studies that 
used minimally coupled massive or massless KG equations 
[14-18,24,25,32,46-51].

\subsection{Time and angular dependent solutions}

After separating the variables $\Phi (t, r, \theta, \phi) =
\Phi_1 (t)\ \Phi_2 (r, \theta, \phi)$ in Eq. (\ref{eq9}), and 
calculating the Laplace-Beltrami operator $\nabla_{\mu} 
\nabla^{\mu}$ for the Schwarzschild metric, the md-KG 
equation becomes
\[
\frac{1}{\Phi_1} \left ( \frac{d^2}{dt^2} + \Omega_o^2 
\right) \Phi_1 = \frac{1}{\Phi_2} \frac{c^2}{r^2} \left ( 1 
- \frac{R_S}{r} \right ) \frac{\partial}{\partial r }  \left [ 
r^2 \left ( 1 - \frac{R_S}{r} \right ) \frac{\partial}
{\partial r} \right ] \Phi_2  
\]
\begin{equation}
\hskip0.05in + \frac{1}{\Phi_2} \frac{c^2}{r^2} \left ( 1 
- \frac{R_S}{r} \right ) \left [ \frac{1}{\sin \theta} 
\frac{\partial} {\partial \theta} \left ( \sin \theta \frac{\partial}
{\partial \theta} \right ) + \frac{1}{\sin^2 \theta} \frac{\partial^2} 
{\partial \phi^2} \right ] \Phi_2 = - \kappa_t^2\ ,
\label{eq11}
\end{equation}
where $\kappa_t^2$ is a time-separation constant, which is assumed 
to be real and positive.  

The resulting differential equation is 
\begin{equation}
\frac{d^2 \Phi_1}{dt^2} = - \left ( \Omega_o^2 + \kappa_t^2 \right ) 
\Phi_1\ .
\label{eq12}
\end{equation}
With the solution $\Phi_1 (t) = \Phi_o \exp{(- i \omega t)}$, where
$\Phi_o$ = const, we obtain $\omega = \pm \sqrt{\Omega_o^2 + 
\kappa_t^2}$.  If $\omega$ is specified, then the separation constant 
can be determined and it is given by  
\begin{equation}
\kappa_t^2 = \omega^2 - \Omega_o^2\ . 
\label{eq13}
\end{equation}
The requirement $\kappa_t^2 \geq 0$, or $\kappa_t$ being real, 
sets up the constraint on $\omega$, which must obey $\omega^2 
\geq \Omega_o^2$.  This guarantees that the temporal solutions 
are oscillatory.  Note that in a special case of $\omega^2 = 
\Omega_o^2$ or $\kappa_t^2 = 0$, the field $\Phi_1 (t)$ 
oscillates with the characteristic frequency $\Omega_o$, which 
after its multiplication by $\hbar$ becomes equal to the rest 
energy of the complex scalar field with mass $m_o$.

As the next step, the field $\Phi_2$ is separated into 
$\Phi_2 (r, \theta, \phi, t) = u(r) v(\theta) w(\phi)$.  By 
taking $w (\phi)$ to be periodic and given as $w (\phi)
= w_o \exp(- i m \phi)$, where $m$ = const, Eq. 
(\ref{eq11}) for $v (\theta)$ can be written as
\begin{equation}
\frac{1}{\sin \theta} \frac{d} {d \theta} \left [ \sin \theta 
\left ( \frac{d v}{d\theta} \right ) \right ] - \left ( \frac{m^2}
{\sin^2 \theta} - \lambda^2 \right ) v = 0\ ,
\label{eq14}
\end{equation}
and its solutions are spherical harmonics $Y_l^m (\theta, \phi)$,
where $l$ and $m$ represent the degree and order of a given 
harmonic, respectively.  In addition, the angular-separation 
constant is defined as $\lambda^2 = l (l+1)$.

\subsection{Radial dependent equation}

Alfter separting and solving the time and angular components 
of the md-KG equation, the remaining radial component for $u(r)$ 
has the following form
\[
\frac{d^2 u}{d r^2} + \frac{1}{r} \left [ 2 + \left ( \frac {R_S}
{r - R_S} \right ) \right ] \frac{du}{dr} +  \left [ \left ( \frac{r}
{R_S} \right )\left ( \frac {R_S}{r - R_S} \right ) \left ( \frac{k_t}
{c} \right ) \right ]^2 u
\]
\begin{equation}
\hskip0.5in - \frac{r}{R_S} \left ( \frac {R_S}{r - R_S} \right ) 
\frac{l (l+1)}{r^2}\ u = 0\ .
\label{eq15}
\end{equation}

Defining $\eta = (r - R_S) / R_S$, which gives $r/R_S = \eta + 1$ 
and $d \eta = (1/R_S) dr$, and taking $u (r) \rightarrow u (\eta)$, 
Eq. (\ref{eq15}) becomes
\begin{equation}
\frac{d^2 u}{d \eta^2} + \frac{2 \eta + 1}{\eta (\eta + 1)} 
\frac{du}{d \eta} +  \left [ \left ( \frac{\eta + 1} {\eta} \right )^2 
\beta_t^2 - \frac{l (l + 1)}{\eta (\eta + 1)} \right ] u = 0\ ,
\label{eq16}
\end{equation}
where $\beta_t = (k_t R_S / c)$.  We find no analytical solutions 
(including the case $l = 0$) given in terms of either elementary 
or known special functions [59,60].  However, it may be possible 
to find new general power series solutions to this equation, which 
is out of scope of this paper.  Instead, our main objective is to 
find solutions to this equation by considerating its simplified 
forms in the following three limited cases: $\eta \rightarrow 
\infty$, $\eta \rightarrow 0$ and $\eta \rightarrow -1$.  
Moreover, all solutions presented below are obtained 
with $l = 0$. 

Before such solutions are presented, we first discuss 
the physical meaning of the dimensionless coefficient 
$\beta_t$, whose explicit form written by using Eq. 
(\ref{eq13}) is 
\begin{equation}
\beta_t^2 = (\omega^2 - \Omega_o^2) t_S^2 =
(\omega t_S)^2 - (\Omega_o t_S)^2\ ,
\label{eq17}
\end{equation}
where $t_S \equiv R_S / c$ is the Schwarzschild time.  
Since $\Omega_o = m_o c^2 / \hbar = c / \bar \lambda_C$, 
with $\bar \lambda_C = \hbar / (m_o c)$ being the reduced 
Compton wavelength, $\Omega_o$ is also known as the 
Compton frequency [6].  According to Eq. (\ref{eq17}),
the Compton frequency multiplied by the Schwarzschild 
time plays an important role in evaluating $\beta_t^2$,
which is required to be either positive or zero (see Eq. 
\ref{eq14}). 

To discuss the physical meaning of the term that contains
the Compton frequency and the Schwarzschild time, we
define $\alpha_{crit} \equiv \Omega_o t_S$, and write
it in the following forms 
\begin{equation}
\alpha_{crit} = R_S\ \bar \lambda_C^{-1} \hskip0.2in
{\rm or} \hskip0.2in 
\alpha_{crit} = 2 \left 
( \frac{m_o}{M_{P}} \right ) \left ( \frac{M}{M_{P}} 
\right )\ , 
\label{eq18}
\end{equation}
where $M_{P} = \sqrt{c \hbar / G}$ is the Planck mass 
(e.g., [36]) that naturally arises in the presented theory.  
Moreover, since $M_P$ is the lower limit for masses of 
classical black holes, and $M_P$ is also the upper limit 
for masses of quantum particles, the gravity effects 
require quantum approach if either $M \sim M_P$ 
or $m_o \sim M_P$ [61].  In this paper, we assume 
that the backreaction of the complex scalar field on 
the metric is negligible, and that all the gravity effects 
on the field are represented by a given metric.  Thus, 
the derived md-KG equation and its solutions are valid 
for black holes with masses $M \gg M_P$, and for 
the fields with masses $m_o \ll M_P$, which 
corresponds to $R_S \gg \bar \lambda_C$. 

\subsection{Solutions to radially dependent equation}

As stated above, the three different limits are applied to 
Eq. (\ref{eq16}), and they correspond to asymptotically 
flat spacetime, the event horizon singularity and the 
central singularity, respectively.  In each case, solutions 
are found to the radially dependent equation that is 
obtained from the md-KG equation in the Schwarzschild 
metric.  The resulting solutions are presented and their 
physical implications on the behavior of the complex 
scalar field in this curved spacetime are discussed. 

\bigskip
\noindent
{\it Case 1}:  In the limit $\eta \rightarrow \infty$, which 
corresponds to $r \rightarrow \infty$, Eq. (\ref{eq16}) 
can be written as 
\begin{equation}
\frac{d^2 u}{d \eta^2} + \frac{2}{\eta} \frac{d u}{d \eta} + 
\beta_t^2 u = 0\ ,
\label{eq19}
\end{equation}
and its solutions are 
\begin{equation}
u (\eta) = C_1 \frac{\e^{- i \beta_t \eta}}{\eta} - i C_2 
\frac{\e^{- i \beta_t \eta}}{2 \beta_t \eta}\ ,
\label{eq20}
\end{equation}
where $C_1$ and $C_2$ are the integration constants, and since $\eta$ 
and $\beta_t$ are real, both solutions are oscillatory with their amplitudes
decreasing to zero when $\eta \rightarrow \infty$.  The same solutions to 
the KG equation in the Minkowski spacetime are obtained when spherical 
coordinates are used [6,47,48].  

\bigskip
\noindent
{\it Case 2}:  Taking the limit $\eta \rightarrow 0$, which corresponds to 
$r \rightarrow R_S$, Eq. (\ref{eq16}) becomes
\begin{equation}
\frac{d^2 u}{d \eta^2} + \frac{1}{\eta} \frac{d u}{d \eta} + 
\left ( \frac{\beta_t}{\eta} \right )^2 u = 0\ ,
\label{eq21}
\end{equation}
and its solutions are 
\begin{equation}
u (\eta) = C_3 \cos( \beta_t \ln \vert \eta \vert ) + C_4 
\sin( \beta_t \ln \vert \eta \vert ) 
\label{eq22}
\end{equation}
where $C_3$ and $C_4$ are the integration constants.  The obtained 
solutions are finite and oscillatory, with their periods getting rapidly 
shorter and shorter when $r$ is approaching $R_S$.  Since Eq. 
(\ref{eq21}) is symmetric with respect to changing $\eta$ to 
$- \eta$, the solutions given by Eq. (\ref{eq22}) are valid on both 
sides of $r = R_S$, but their validity is restricted only to the vicinity 
of the event horizon.  However, according to Eq. (\ref{eq22}), the 
solutions are undefined at $\eta = 0$ or $r = R_S$; for the value 
of $\Phi (R_S)$ see Section 4.2.

\bigskip
\noindent
{\it Case 3}:  In order to consider the limit $r \rightarrow 0$,
we apply this limit to  Eq. (\ref{eq15}), and introduce $\xi = 
r / R_S$.  The resulting equation is 
\begin{equation}
\frac{d^2 u}{d \xi^2} + \frac{1}{\xi} \frac{d u}{d \xi} + 
\xi^2 \beta_t^2 u = 0\ ,
\label{eq23}
\end{equation}
with its solutions given by  
\begin{equation}
u (\xi) = C_5\ J_o \left ( \frac{1}{2} \beta \xi^2 \right ) +
C_6\ Y_o \left ( \frac{1}{2} \beta \xi^2 \right )\ , 
\label{eq24}
\end{equation}
where $C_5$ and $C_6$ are the integration constants, and 
$J_o (\beta_t \xi^2 / 2)$ and $Y_o (\beta_t \xi^2 / 2)$ are 
Bessel functions of the first and second kind, respectively. 
Note that as $\xi \rightarrow 0$, the solutions $J_o (\beta_t 
\xi^2 / 2) \rightarrow 1$ and $Y_o (\beta_t \xi^2 / 2) 
\rightarrow - \infty$.  Since the second solution is unphysical,
we set $C_6 = 0$.  With $J_o (\beta_t \xi^2 / 2)$ being the 
only physically acceptable solution, we set $\xi = 0$ and write
the solution as $u (0) = C_5$.  This shows that the solution 
at the central singularity ($\xi = 0$ or $r = 0$) is given by 
a constant $C_5$, which is real and finite (see Section 4.2).

The above studies of the solutions to the radially dependent 
equation of the complex scalar field $u (r)$ demonstrate 
that the field is oscillatory in asymptotically flat spacetime,
and that it remains oscillatory down to the black hole's 
event horizon.  The solutions are also oscillatory in the 
vicinity of the event horizon and the period of these 
oscillations rapidly decreases when $r$ approaches $R_S$ 
from the outside of the event horizon.  Similar behavior 
of the field is observed when $r \rightarrow R_S$ with 
its negative values from the black hole's interior.  In 
other words, the field $u(r)$ is oscillatory on both sides 
of the event horizon.  However, at $r = R_S$, the radial
component of the scalar field cannot be evaluated 
mathematically because its limit does not exist.  On 
the other hand, the oscillatory behavior of the field 
in time, $\Phi_1 (t)$, is not affected by this singularity.

The obtained results also show that in the vicinity of the 
central singularity the radial component of the field $u(r)$ 
is oscillatory and that its value is constant at $r = 0$. Thus, 
the field is not affected by the presence of this black hole 
singularity.  Similarly, this singularity has no effect on 
the oscillatory behavior of the field in time.  Thus, we 
conclude that the radial component of the complex scalar 
field described by the md-KG equation is oscillatory on 
the Schwarzschild spacetime except at the event horizon, 
where its value cannot be determined using the approximate 
solution, and at the central singularity, where it reaches its 
constant value.  For discussion of the behavior of $\Phi 
(t, r, \theta, \phi)$ in the entire Schwarzschild spacetime 
see Section 4.2. 

\subsection{Metric-dependent Klein-Gordon equation in 
Lemaitre coordinates}
 
We now introduce the Lemaitre coordinates (e.g., [34-37]), 
with $x = (cT, R, \theta, \phi))$ and use Eq. (\ref{eq6}) 
to obtain the md-KG equation in the following form 
\begin{equation}
\left [ \nabla_{\mu} \nabla^{\mu} + \left ( 1 - \frac{R_S}
{r (R, cT)} \right )^{-1} \Omega_o^2 \right ] \Phi (x_L) = 0\ ,
\label{eq25}
\end{equation}
which shows that the resulting equation is of the same form as 
the md-KG equation for the spherical coordinates $x_S$ (see Eq. 
\ref{eq8}).  This means that Eq. (\ref{eq25}) is diffeomorphism 
invariant.   There is a relationship between the coordinates 
$x_L$ and $x_S$, specifically, the variable $r$ is expressed in 
terms of $cT$ and $R$, and given by 
\begin{equation}
r (R, cT) = \left [\frac{3}{2} \sqrt{R_S} ( R - cT ) \right ]^{2/3}\ .
\label{eq26}
\end{equation}
It is seen that the second term in the md-KG equation changes 
sign when $r (R, cT) > R_S$, which gives different solutions 
than those obtained for $r (R, cT) < R_S$.  Similar to Eq. 
(\ref{eq9}), the md-KG equation in the Lemaitre coordinates 
has its singularity at $r(R, cT) = R_S$, which corresponds 
to $(R - cT) = (2/3) R_S$.  However, the equation shows 
no singularity at $r(R, cT) = 0$.  An interesting result is 
that despite the fact that the Lemaitre coordinates remove 
the singularity at $r = R_S$ from the Schwarzschild metric, 
the black hole's event horizon remains a singular point in 
the md-KG equation that describes the field $\Phi (x)$. 

Since Eqs (\ref{eq25}) and (\ref{eq10}) are of the same form,
they are diffeomorphism invariant.  Thus, the values of the 
field $\Phi$ near the black hole singularities remain the same 
in any coordinate system.  The results are easy to verify by 
finding solutions to Eq. (\ref{eq25}) for the same cases as 
those discussed in Section 3.  This shows that diffeomorphism
invariance guarantees that the results obtained in spherical and 
Lemaitre coordinates are consistent for both falling and distant
observers.  However, these observers see different pictures of 
the field near and at the event horizon because the falling 
observer reaches and crosses the horizon in a finite amount 
of proper time (for details, see Section 4.5).

\subsection{Comparison to previous solutions}

In previous work based on the KG equation $[ \nabla_{\mu} 
\nabla^{\mu} + m_o^2] \Phi = 0$, some series solutions 
to this equation were found in the Schwarzschild spacetime 
[48,49], and they demonstrated that the solutions were finite 
at the central singularity and everywhere at $r > R_S$.  However,
their validity in the vicinity of $r = R_S$ could not be uniquely 
established.  The results presented in this paper show that the 
field $\Phi$ is an oscillatory function of $r$ in the vicinity of
$r = R_S$, and that it becomes zero at this singularity.  Attempts
to quantize scalar fields described by the KG equation inside the 
Schwarzschild black hole were done in [50,51], and more recently 
in [47], who used the T-model for the interior variables [46].
The main result of these papers is that the proposed quantization 
procedures break down near and at the central singularity. 
However, the solutions to the md-KG equation inside the black 
hole, and specifically near and at the central singularity show
that $\Phi$ is a well-behaving function when $r \rightarrow 0$,
and it reaches its zero value at $r = 0$. 

\section{Physical implications of the obtained results}

\subsection{Comparison to previously used Klein-Gordon equations}

The derived md-KG equation (see Eq. \ref{eq6}) is valid for 
curved spacetimes represented by a 4D pseudo-Riemannian 
manifold.  The equation is derived from the local eigenvalue 
equation for the energy-momentum operator (see Eq. \ref{eq3}) 
with its eigenvalues $p_{\mu} (x)$ uniquely determined by 
a given metric.  For flat spacetime with the Minkowski metric, 
the term $p_{\mu} (x) p^{\mu} (x) = \Omega_o^2 = m_o^2 
c^4 / \hbar^2$ = constant.  This shows that the presence of 
the tensor field $p_{\mu} (x) \neq 0$ becomes the source of 
mass in the md-KG equation for the complex scalar field $\Phi (x)$; 
note that $p_{\mu} (x) = 0$ makes the field massless.  Thus, 
the origin of mass for the field $\Phi (x)$ is caused by the 
presence of $p_{\mu} (x)$, and no other mass generation 
mechanism is needed. 

The md-KG equation significantly differs from the KG equation 
that was used in previous studies of scalar fields in curved 
spacetime (e.g., [14-18]).  The commonly used KG equation 
in the past was  $[ \nabla_{\mu} \nabla^{\mu} + \xi R (x) 
+ m_o^2] \Phi = 0$, where $\xi$ is a dimensionless constant, 
$R (x)$ is the Ricci scalar curvature of a given spacetime, and 
$m_o = \Omega_o$ = const represents the rest mass in natural 
units.  The term $\xi R (x)$ was introduced by imposing conformal 
invariance on the KG equation [14,16-18], and this term is assumed 
to describe a coupling between the scalar and gravitional fields.  
The KG equation with this term was used to formulate QFT in 
globally hyperbolic spacetimes [14-18].  

In the Schwarzschid metric, $R(x) = 0$, and the KG equation 
becomes $[ \nabla_{\mu} \nabla^{\mu} + m_o^2] \Phi = 0$
for massive (e.g., [30,31,33,41-46]) or massless, with $m_o = 
0$, (e.g., [18,24,25,32]) fields.  The energy-momentum 
relationship $E^2 = p^2 c^2 + m_o^2 c^4$, where $E$ and 
$p$ are eigenvalues of the energy and momentum operators, 
respectively, and $m_o$ = const, underlies the KG equation 
in the Minkowski spacetime.  In the previous work cited above, 
it was assumed that this relationship remains globally valid in 
the entire Schwarzschild spacetime.  However, the md-KG equation 
given by Eq. (\ref{eq9}) significantly differs from the previously 
used KG equations because of the presence of the factor $(1 - 
R_S /r)^{-1}$.  This factor shows that in the Schwarzschild 
spacetime the energy-momentum relationship is valid only locally, 
and that its correct global form can only be obtained from the term 
$p^{\mu} (x) p_{\mu} (x)$.  Note also that the factor significantly 
alters physical description of the scalar field in this spacetime as 
shown by the solutions presented above.  Moreover, the form of 
the md-KG equation is metric dependent, which means that the form 
of the term $p^{\mu} (x) p_{\mu} (x)$ will be different on different 
metric manifolds.

The md-KG equation (see Eq. \ref{eq10}) was used to investigate 
the complex scalar field in the vicinity of, and at the event horizon, 
as well as near and at the central singularity of a Schwarzschild 
black hole.  The solutions were obtained for both spherical and 
Lemaitre coordinates in the Schwarzschild metric, and they 
show that at both these singularities the field remains finite, 
which means that it is singularity-free.  The obtained results 
also demonstrate that the event horizon separates the field 
inside the black hole from the one outside, and that the 
behavior of these to two fields is very different as it is 
now discussed. 

\subsection{Potential for scalar field and its extreme values}

The standard Lagrangian for the md-KG equation given by Eq. 
(\ref{eq10}) can be written as
\begin{equation}
{\cal L}_{\phi} = ( \nabla_{\mu} \Phi ) ( \nabla^{\mu} \Phi ) 
- V (r, \Phi)\ , 
\label{eq27}
\end{equation}
where $\Phi = \Phi (t, r, \theta, \phi)$ and the potential 
$V (r, \Phi)$ is given by 
\begin{equation}
V (r, \Phi) = \frac {1}{2} \left ( 1 - \frac{R_S}{r} \right )^{-1} 
\Omega_o^2 \Phi^2\ . 
\label{eq28}
\end{equation}
The md-KG equation has a singular point at $r = R_S$, as 
does the potential.  As a result, the potential changes its 
sign at the event horizon and becomes $V (r, \Phi) > 0$ for 
all $r > R_S$, which indicates stability of the field [14,36].
However, in the black hole's interior $r < R_S$, the potential 
is $V (r, \Phi) < 0$, which implies its tachyonic character (e.g., 
[52-56]); note that $V (r, \Phi)$ is different from 'tachyonic 
potentials' that arise in string theories (e.g., [57,58]).  

From Eq. (\ref{eq28}), we consider the extrema of $V(r, \Phi(r))$
by taking 
\begin{equation}
\frac{dV}{dr} = \frac{\partial V}{\partial r} + \frac{\partial V}
{\partial \Phi} \frac{d \Phi}{dr} = 0,
\label{eq29}
\end{equation}
which, after some algebra, gives
\begin{equation}
\Phi^2 = \left( \frac{r}{R_s} \right) (r-R_s) \frac{d \Phi^2}{dr}.
\label{eq30}
\end{equation}
The RHS is zero for $r = 0$, or $r = R_s$, which requires 
$\Phi = 0$ at both of these locations.  Now, if $d \Phi^2 / 
dr = 0$, then again only $\Phi = 0$ satisfies the condition. 
According to the results of Section 3.2, $\Phi (t, r, \theta, 
\phi) = \Phi_1 (t) u(r) v( \theta ) w (\phi)$, with the solutions 
for $\Phi_1 (t)$, $v( \theta )$ and $w (\phi)$ being global 
solutions valid in the entire Schwarzschild spacetime.  
However, there are no such global solutions for $u(r)$ 
but only approximate solutions valid in the vicinity of 
the singularities at $r = 0$ and $r = R_s$.

Because of this lack of global solutions for $u(r)$, we cannot 
uniquely determine reasons for $\Phi (t, R_S, \theta, \phi) = 0$,
because the approximate solution for $u(R_S)$ is mathematically 
undefined (see Eq. \ref{eq22}).  Based on this limited information, 
we may conclude that $\Phi (t, R_S, \theta, \phi) = 0$ is satisfied 
only if $\Phi_1 (t)$ has a node at $r = R_S$.  Similarly, $\Phi_1 (t)$
must also have a node at $r = 0$ to satisfy $\Phi (t, 0, \theta, \phi) 
= 0$, since $u (0) = C_5$ (see Eq. \ref{eq24}).     

Using the above results, we calculate the forces resulting from the 
potential and obtain 
\begin{equation}
F_r (r, \Phi) = - \frac{\partial V}{\partial r} = \frac {1}{2} \left (
\frac{1}{r - R_S} \right )^2 R_S\ \Omega_o^2\ \Phi^2\ , 
\label{eq31}
\end{equation}
and 
\begin{equation}
F_{\Phi} (r, \Phi) = - \frac{\partial V}{\partial \Phi} = - \left (
\frac{r}{r - R_S} \right ) \Omega_o^2\ \Phi\ .
\label{eq32}
\end{equation}
The positive force $F_r (r, \Phi)$ acts on the field along the 
$r$-direction.  However, the force $F_{\Phi} (r, \Phi)$ for 
a fixed value of $r$ is similar to that acting in a linear 
harmonic oscillator.  This means that $F_{\Phi} (r, \Phi)$
is the restoring force that pushes the field back toward the 
minimum.  This becomes important in our studies of the 
field outside and inside of the black hole that is presented 
below. 

By substituting Eq. (\ref{eq30}) into Eq. (\ref{eq31}) 
and using the l'Hopital rule, we obtain $F_r (R_S, 
\Phi (R_S)) = 0$, which shows no force at the event 
horizon in the $r-$direction.  Similarly, using Eqs   
(\ref{eq30}) and (\ref{eq33}), we find $F_{\Phi} 
(0, \Phi (0)) = 0$, so no force in the $r-$direction 
at the central singularity either.  Note that these 
forces result from the potential of the standard 
Lagrangian for the md-KG equation.  

The results demonstrate that there are are two stable 
minima, one at $V_{min}^{out} (R_S, \Phi (R_S)) = 0$ 
for the oustide field, and one at $V_{min}^{in} (0, \Phi
(0)) = 0$ for the field inside the black hole.  At both 
minima the field $\Phi = 0$, which means that they 
represent two stable vacuums for the corresponding 
scalar fields. 

\subsection{Complex scalar field outside the black hole}

We normalize the potential by taking $V (r_o, \Phi) / 
\Omega_o^2 = (1/2) [r_o / (r_o - R_S)] \Phi^2$, with 
$r_o$ being fixed, and plot it in Fig. 1 as a function of 
$\Phi$ for $r_o > R_S$.  The plots demonstrates that 
the potential has rotational symmetry and that its 
minimum at $r = R_S$ is the same for the field for 
all the values $r_o$ outside the event horizon.  As 
shown above, this minimum represents a vacuum for 
the outside field and this vacuum is stable.  Since the 
plotted potential is similar to the harmonic oscillator 
potential, the field oscillates with respects to its minimum 
value, and the range of these oscillations for a given $r_o$ 
is determined by the width of each parabolic curve in Fig. 1.  
Note that the minimum $V (\Phi) = 0$ also represents stable 
vacuum in the asymptotically flat (Minkowski) spacetime.  
%

%
\begin{figure}[ht]
\begin{center}
\includegraphics[width=125mm]{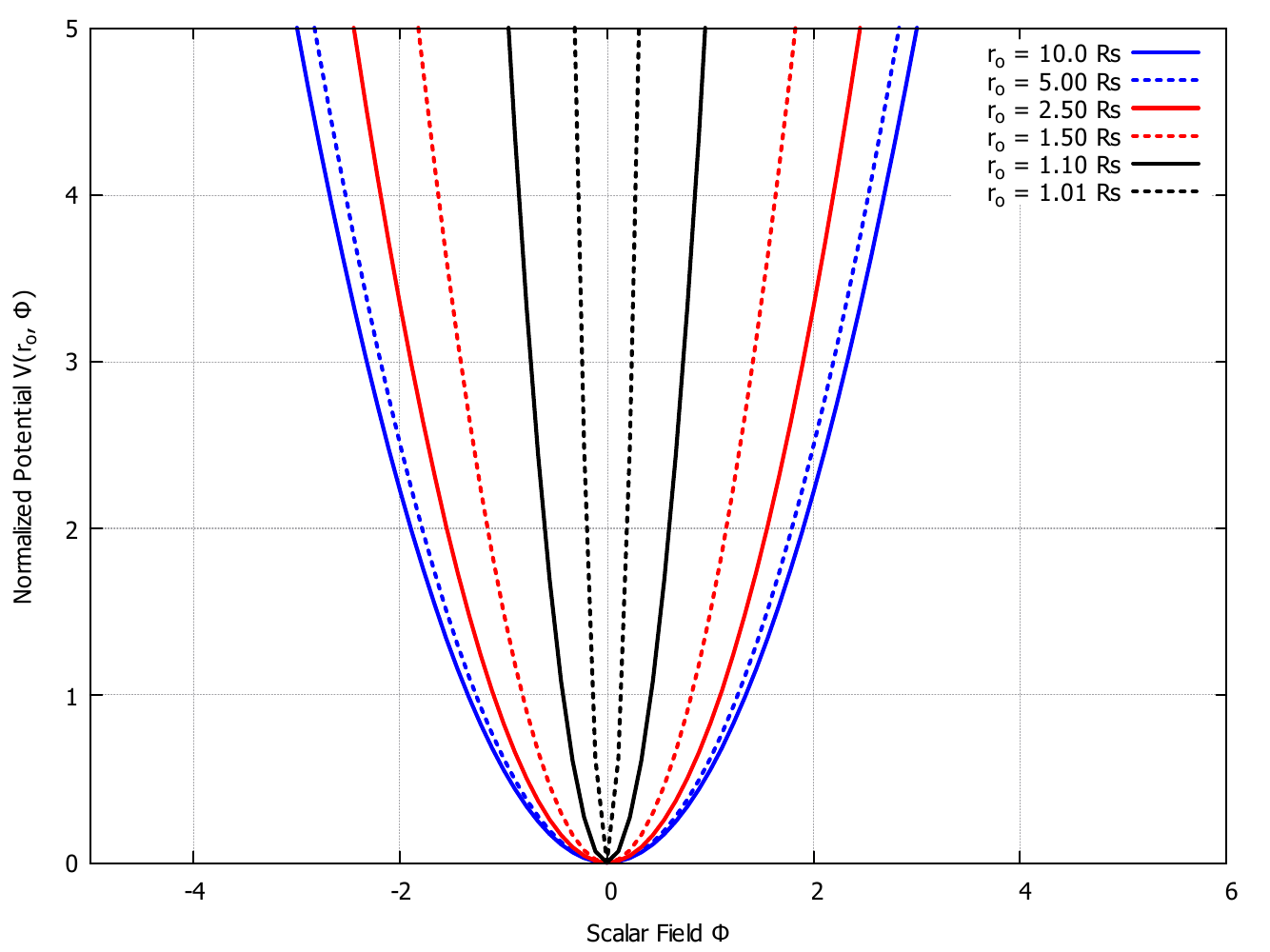}
\caption{The normalized potential $V (r_o, \Phi) / \Omega^2$ 
is plotted as a function of the outside field $\Phi$ for several 
fixed values of $r_o > R_S$.}
\label{fig.1}
\end{center}
\end{figure}
%

According to the results of Section 4.2, the potential $V 
(R_S, \Phi (R_S)) = 0$.  However, its value increases in 
the vicinity of the event horizon if $r = R_S + \epsilon$ 
for small $\epsilon$ (see Eq. \ref{eq28}).  The result of 
this potential increase is formation of a potential barrier 
near the event horizon for the complex scalar field.  The 
barrier prevents the outside field from entering the black 
hole and causes formation of a scalar hair.  The resulting 
scalar hair is a stable feature around the black hole.  
Because of the different nature of the potential in the 
md-KG equation, the origin of this scalar hair is very 
different than those obtained in [30-33], who used 
the minimally coupled massive KG equation or limited 
their studies to the massless KG equation.   In previous 
work, scalar hair was typically generated by scalar 
potentials of specific forms that produced the desired 
results, or by specifying an initial periodically time-varying 
but spatially homogeneous scalar background [31].

In the approach presented in this paper, the distant observer 
in spherical coordinates sees a long-lived hair around the black 
hole formed by the complex scalar field.  Radiation from this 
field is subjected to gravitational redshift, which causes the 
field's freezing at the horizon for the distant observer [36].
However, the falling observer experiences a peak in energy 
density near the event horizon, and this peak can be detected 
by this observer, who reaches and crosses the horizon in a 
finite amount of proper time [34-36]. 

\subsection{Tachyonic field inside the black hole}

In the Schwarzschild black hole's interior, the term $(1 - 
R_S/r)^{-1} \Omega_o^2$ in the md-KG equation (see 
Eq. \ref{eq10}) changes its sign, which makes the field 
$\Phi$ resulting from this equation to be tachyonic 
(e.g., [3,4]).  Note that the origin of this tachyonic field 
is caused by changes in the metric at the event horizon, 
which is more physical than the requirement that 
$m_o^2 < 0$ postulated in some previous studies 
(e.g., [52-54] and references therein).  Another 
well-known example of tachyonic field is the Higgs 
field, which is initially in an unstable state caused 
by its self-coupling properties.  The unstable field 
decays rapidly into a stable state, and as a result 
one massive and one massless field are generated 
(e.g., [3,4]).  This process of field decay is known 
as a tachyon condensation (e.g., [55,56]).  Since 
the field $\Phi$ considered here is not self-interacting, 
the origin of its tachyonic nature and its behavior 
are very different from that observed in the Higgs 
mechanism. 
%

%
\begin{figure}[ht]
\begin{center}
\includegraphics[width=125mm]{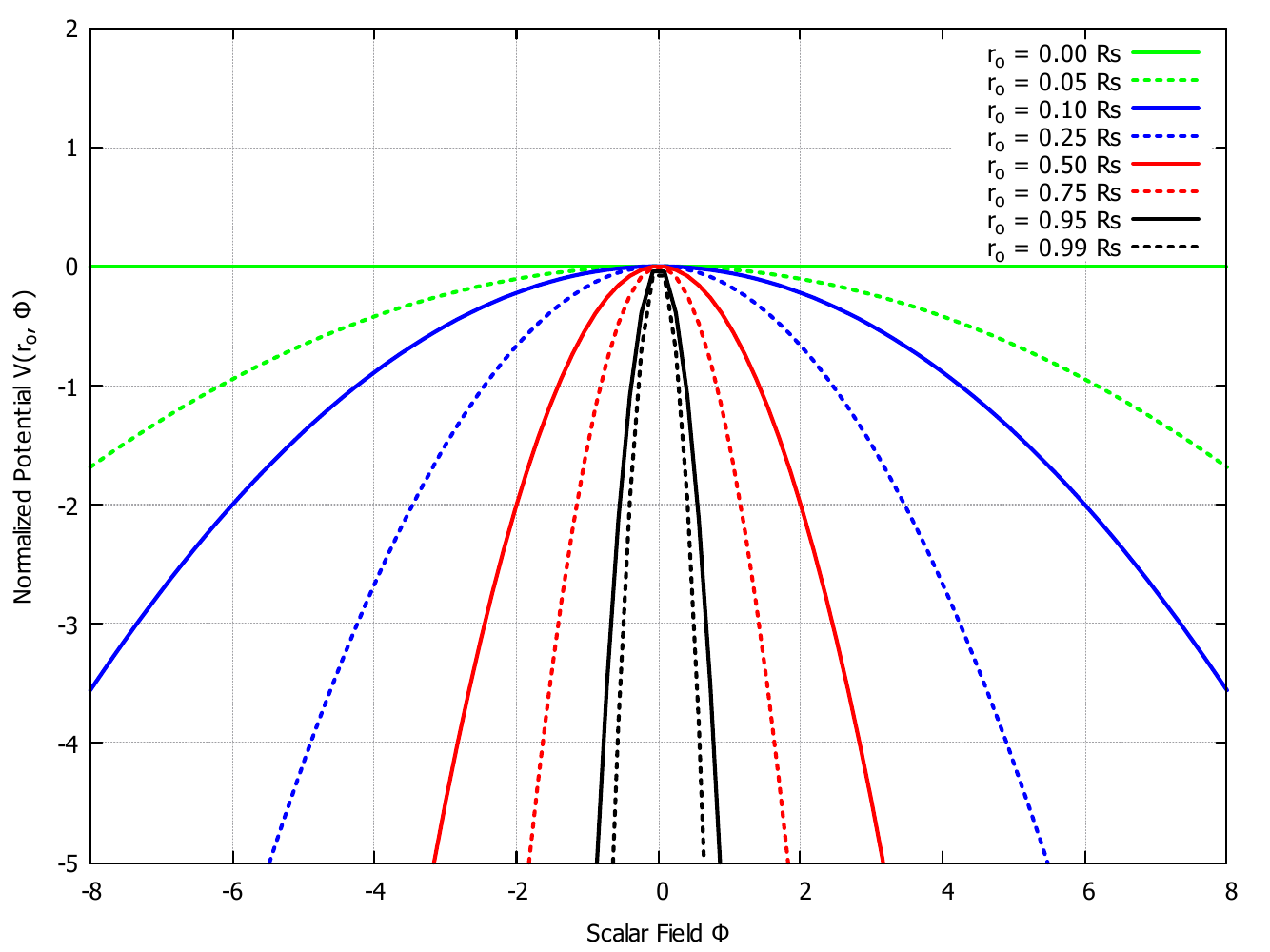}
\caption{The normalized potential $V (r, \Phi) / \Omega^2$ 
is plotted as a function of the inside field $\Phi$ for several 
fixed values of $r > R_S$.}
\label{fig.1}
\end{center}
\end{figure}
%

The normalized potential $V (r_o, \Phi) / \Omega^2$ is 
plotted in Fig. 2 as a function of $\Phi$ for several fixed 
values of $r_o < R_S$, which means that the plots are for 
the inside field only.  The plots show that the potential 
has rotational symmetry and that its maximum is at $r 
= R_S$, where $V (R_S, \Phi (R_S)) = 0$.  Because the 
field is tachyonic, the maximum is unstable.  An interesting
result is that $V (R_S, \Phi (R_S)) = 0$ represents the 
stable minimum for the outside field and it is also the 
unstable maximum for the inside field. 

By being tachyonic, the inside field rapidly decays into its 
stable configuration that is represented by $V (0, \Phi (0)) 
= 0$, which is the location of stable vacuum for the field. 
The tachyonic field decays by rolling down from its unstable 
maximum, located at $r = R_S$, to stable minimum located 
at $r = 0$.  This process is called the tachyonic condensation 
and the rolling down of the field takes place along the $r$ 
coordinate that plays the role of the temporal-like coordinate 
inside the black hole [36,47].  During the rolling down process, 
energy is released and this energy is responsible for converting 
the tachyonic field into a scalar field with a positive potential.  
This is in agreement with the solutions obtained to the asymptotic 
forms of the md-KG equations that were derived on both sides 
of the event horizon (see Eq. \ref{eq22} and discussion below
this equation), and in the vicinity of the central singularity 
(see Eq. \ref{eq24} and its relevant discussion). 

The process of tachyonic condensation described above 
is different from that in the Higgs mechanism, where the 
condensation is responsible for generation of one massive 
and one massless field.  Our results show that the rest 
mass $m_o$ of $\Phi$ remains unchanged because 
$\Omega_o = m_o c^2 / \hbar$ is the same throughout 
the entire outside and inside spacetime of the black hole.  
Therefore, the energy resulting from the tachyonic 
condensation inside the black hole is used to convert 
the original tachyonic field into the field that is similar 
to that existing outside of the black hole.  

Thus, after the tachyonic condensation inside the black 
hole, the outside and inside scalar fields are described 
by the same md-KG equation whose potential is repulsive, 
$V (r, \Phi) > 0$.  However, the potential of the tachyonic 
field is attractive, $V (r, \Phi) > 0$ before its condensation
into a stable repulsive field.  Despite the fact that both 
inside and outside fields are stable and repulsive, there 
is a fundamental difference between them, namely, each 
field has its different minimum, which corresponds to its 
stable vacuum.  For the outside field, the stable minimum 
$\Phi = 0$ is at $r = R_S$, but for the inside field, the 
stable minimum $\Phi = 0$ is at $r = 0$ and is associated 
with the central singularity of the black hole.  The shows 
that the outside and inside fields are very different.

Based on the above results, it is seen that the origin of 
the tachyonic field is caused by the sign change in the 
Schwarzschild metric when the event horizon is crossed.  
Thus, this change is responsible for the tachyonic 
condensation and its effects on the field inside the 
black hole.  This makes the described mechanism 
different from that originally proposed by Higgs.  

\subsection{Classical observers and measurements}

In spherical coordinates of the Schwarzschild metric, 
stationary classical observers can be defined either at very 
large distances from the black hole $r \rightarrow \infty$ 
or on stable orbits around the black hole, and they are 
called 'shell-observers' [36].  However, in the Lemaitre
coordinates, all classical observers are falling along 
radial geodesics and their frames are co-moving since
they record the same proper times [36].  In the following,
we consider which physical quantities can be measured 
by these observers and what will be the results of such 
measurements.

Let $\Omega_{eff}^2 (r) \equiv ( 1 - R_S / r )^{-1}\ \Omega_o^2$ 
be the field's effective characteristic frequency, and the effective 
mass of the field be $m_{eff}^2 (r) \equiv ( 1 - R_S / r )^{-1}\ 
m_o^2$.  Then, $\Omega_{eff}^2 (r) = m_{eff}^2 (r) c^4 / 
\hbar^2$.  A distant stationary observer located far away from the 
black hole, where spacetime is asymptotically flat, sees $m_{eff} (r)$ 
rapidly increasing to infinity as $r \rightarrow R_S$.  If there is a
particle of mass $m_o$ associated with the field at the observer's 
location, and this particle is falling from rest at infinity radially onto 
the black hole, then its velocity $v_{d} (r)$ as seen by this distant 
observer [36], who now uses Eq. (\ref{eq10}), is given by 
\begin{equation}
v_{d} (r) = - c \sqrt{\frac{R_S}{r}} \frac{m_o^2}{m_{eff}^2 (r)}\  .
\label{eq33}
\end{equation}
This shows that if $r \rightarrow R_S$, then $m_{eff} (r) \rightarrow 
\infty$ and $v_{d} (r) \rightarrow 0$ as seen by the distant observer.  
Note also that at $r \rightarrow \infty$, the effective mass $m_{eff} 
(r) \rightarrow m_o$ and the velocity $v_{d} (r) \rightarrow 0$.  
Now, if the particle is falling from rest at infinity, its velocity relative 
to a stationary observer at a radial distance $r$ is commonly known 
as the 'shell velocity' $v_{s} (r)$ [36], and according to Eq. (\ref{eq10}), 
it can be written in the follwoing form
\begin{equation}
v_{s} (r) = \pm c \sqrt{1 - \frac{m_o^2}{m_{eff}^2 (r)}}\  .
\label{eq34}
\end{equation}
In the limit of $r \rightarrow R_S$, $m_{eff} (r) \rightarrow 
\infty$ and $v_{s} (r) \rightarrow \pm c$.  However, when 
$r \rightarrow \infty$, then $m_{eff} (r) \rightarrow m_o$ 
and $v_{s} (r) = v_{d} (r) \rightarrow 0$.  This shows that
both the distant and shell observers can only measure in 
their stationary frames the value of $m_o$, which is then 
used to determine $\Omega_o$.  In other words, their 
measurements are always independent of the variables 
$t$, $r$, $\theta$ and $\phi$, which are not directly 
measurable coordinates. 

Since the results given by Eqs (\ref{eq33}) and (\ref{eq34}) 
are obtained using Eq. (\ref{eq10}), and since they are 
consistent with those found directly from the Schwarzschild 
metric for a classical particle (e.g. [36]), this consistency is 
an independent validation of the derived md-KG equation 
in this metric.  Note also that both the stationary distant 
observer and the shell-observer have no physical access 
to the black hole's interior, which can only be studied 
by these observers theoretically by solving the md-KG 
equation.  However, the situation is different for falling 
observers in the Lemaitre coordinates as is now shown.

In Lemaitre coordinates, observers are in free-fall along 
radial geodesics and they are co-moving, which means 
that the variable $T$ is the proper time in all the co-moving 
frames and that $dR / dT = 0$ in those frames.  The 
falling observers approach the black hole with the local 
escape velocity $v_{e} (r) = dr / dT = c \sqrt{R_S / r}$,
which gives $v_{e} (r = R_S) = c$ [36].  However, this 
does not mean that the falling observers exceed $c$ in 
the black hole interior.  On the contrary, their local 
measurements never show velocities higher than $c$.  
The effect $v_{e} > c$ for $r < R_S$ appears only
to the stationary observers located outside the event 
horizon because space and time roles change at $r = 
R_S$, which means that $dr / dT$ represents the 
flow towards the central singularity, not a physical 
speed that exceeds $c$.  

Since the frames of these observers are in free-fall 
along riadal geodesics, the observers can only 
measure the values of $m_o$ in their falling frames.
Then, they can use their measurements to determine 
the value of $\Omega_o$, which is the characteristic 
frequency of mass $m_o$ of the complex scalar field 
$\Phi$.  Note that $\Omega_o$ = const remains in 
the entire Schwarzschild spacetime, including its 
central singularity.     

\section{Conclusions}

We derived the metric-depedent Klein-Gordon equation for 
a massive and complex scalar field in curved spacetime of 
a given metric using gauge and group theories.  Then, we 
used the equation to study the behavior of the scalar field
outside and inside of  a Schwarzschild black hole.  We 
found the asymptotic solutions to this equation in the 
vicinity of the event horizon and central singularity.  
The obtained solutions are extrapolated to determine 
the field at the locations of both singularities.  Our 
studies of the potential of the complex scalar field 
revealed fundamental differences in the behavior 
of the field outside and inside of the black hole.\\

\noindent
The main obtained results are:\\
(i) The field is oscillatory outside the event horizon, and 
the period of these oscillations rapidly decreases when $r$ 
approaches $R_S$.  At the location of the event horizon, 
the outside field becomes exactly zero, which is its unique 
and stable minimum.\\
(ii) The value of the potential of the outside scalar field in 
the vicinity of the horizon increases, which leads to the 
formation of a stable scalar hair.\\
(iii) The inside scalar field is tachyonic and its maximum,
which is the field-zero-value at the event horizon, is unstable.
This tachyonic field rapidly decays (tachyonic condensation) 
by rolling down along the $r-$coordinate to its true minimum
that is located at the central singularity.\\
(iv) The field resulting from the tachyonic condensation has 
the same rest mass as the outside field and its potential is 
positive, so the field becomes repulsive.  The field does not 
collapse to the central singularity, but instead its forms a
stable standing wave pattern between $r = R_S$ and 
$r = 0$.\\
(v) The tachyonic condensation in black holes is different 
from the one that operates in the Higgs mechanism. The
main physical erason for this complex scalar field behavior
is the change in sign in the Schwarzschild metric at the 
event horizon.\\

Einstein's GR, which is a classical theory, predicts the 
gravitational forces to be infinitely strong at the center 
of black holes, which results in the physical parameters, 
such as density and pressure to become infinitely large 
at this central singularity.  The solutions to the md-KG 
equation for the complex scalar field obtained in the 
vicinity and at the central singularity do not confirm 
these results.  Moreover, the presented results are also
very different than those obtained by the Klein-Gordon 
equation minimally coupled to gravity.  Finally, the results 
of this paper have important physical implications on 
formulation of quantum field theories around and inside 
black holes; however, such theories will be developed 
in a separate paper.\\

\bigskip\noindent
{\bf Acknowledgment}\\
We are grateful to L.D. Swift for many stimulating 
discussions on physics of curved spacetimes and 
related problems.  Our special thanks to staff of 
the French Cafe La Madeleine in Arlington, Texas,
 for allowing us to spend many hours of discussions 
that led to the results of this paper.

\bigskip\noindent
{\bf References}\\  
%


\begin{thebibliography}{qqq}

\bibitem{1} O. Klein, Z. Phys. 37 (1926) 895
\bibitem{2} W. Gordon, Zeits. f\"ur Phys. 40 (1926) 117; 40 (1926) 121
\bibitem{3} M. Kaku, Quantum Field Theory, A Modern Introduction, Oxford
                  Uni. Press, Oxford, 1993
\bibitem{4} P.H. Frampton, Gauge Field Theories, John Wiley \& Sons, Inc., 
                  New York, 2000 
\bibitem{5} H.C. Ohanian, Special Relativity: A Modern Introduction, Phys.
                  Curr. \& Instruct., Lakeville, MN, 2001 
\bibitem{6} W. Greiner, Relativistic Quantum Mechanics, Springer-Verlag, Berlin, 1990
\bibitem{7} W.I. Fushchich and A.G. Nikitin, Symmetries of Equations 
                  of Quantum Mechanics, Allerton Press, Allerton, NY, 1994
\bibitem{8} E.P. Wigner, Ann. Math. 40 (1939) 149
\bibitem{9} Y.S. Kim and M.E. Noz, Theory and Applications of the 
                  Poincar\'e Group, Reidel, Dordrecht, 1986
\bibitem{10} E. In\"onu and E.P. Wigner, Nuovo Cimiento 9 (1952) 705  
\bibitem{11} J.L. Fry and Z.E. Musielak, Ann. Phys. 325 (2010) 2668
\bibitem{12} J.L. Fry, Z.E. Musielak and Trei-wen Chang, Ann. Phys. 326 (2011) 1972
\bibitem{13} Z.E. Musielak, J.L. Fry and G.W. Kanan, Adv. Theor. Phys. 9 (2015) 213
\bibitem{14} R.M. Wald, General Relativity, The University of Chicago Press, 
                    Chicago and London, 1984
\bibitem{15} N.D. Birrell and P.C.W. Davis, Quantum Fields in Curved Space, 
                   Cambridge University Press, Cambridge, 1982
\bibitem{16} S.A. Fulling, Aspects of Quantum Field Theory in Curved spacetime, 
                   Cambridge University Press, Cambridge, 1989
\bibitem{17} R.M. Wald, Aspects of Quantum Field Theory in Curved Spacetime and
                    Black Hole Thermodynamics, The University of Chicago Press, Chicago 
                    and London, 1994
\bibitem{18} L. Parker and D. Toms, Quantum Field Theory in Curved Spacetime, 
                   Quantized Fields and Gravity, Cambridge University Press, Cambridge, 2009
\bibitem{19} M. Apostol, Progress in Physics 1 (2008) 90
\bibitem{20} F. Michael, arXiv:1004.1543v1 [cond-mat.stat-mech] 12 June 2023
\bibitem{21} R.D. Lehn, S.S. Chabysheva and J.R. Hiller, Eur. J. Phys. 39 (2018) 045405
\bibitem{22} G. Nash, Int. J. Mod. Phys. A 36 (2021) 2150196
\bibitem{23} D.A. Taylor, S.S. Chabysheva and J.R. Hiller, arXiv:2212.14166v2 [hep-th] 
                   12 June 2023
\bibitem{24} S.W. Hawking, Commun. Math. Phys. 43 (1975) 199
\bibitem{25} H. Yumisaki, Prog. Theor. Exp. Phys. (2017) 063B04 (43 pages) 
\bibitem{26} J.L. Fry and Z.E. Musielak, Modern Phys. Lett. A, submitted (2026)\\
                    Preprint, DOI: 10.13140/RG.2.2.32305.03686 (March 2026)
\bibitem{27} R. Geroch, Comm. Math. Phys. 26 (1972) 271
\bibitem{28} M. Heller, P. Multarzynski and W. Sasin, Acta Cosmologica 16 (1989) 53
\bibitem{29} M. Heller, Introduction to the Global Structure of spacetime:
                   Theoretical Foundations of Cosmology, World Scientific, Singapore, 1992
\bibitem{30} C. Gao and J. Qiu, Gen. Relat. \& Grav. 54 (2022) 158
\bibitem{31} K. Clough, P.G. Ferreira and M. Lagos, Phys. Rev. D 100 (2019) 063014
\bibitem{32} Z.-H. Yang, C. Xu, X.-M. Kuang, B. Wang, R.-H. Yue, Phys. Lett. B 853 
                   (2024) 138688
\bibitem{33} Y. Brihaye, B. Hartmann and K. Horton, Phys. Rev. D 112 (2025) 044044 
\bibitem{34} S. Weinberg, Gravitation and Cosmology: Principles and Applications of 
                   the General Theory of Relativity, Wiley, New York, 1972           
\bibitem{35} J.A. Wheeler, C. Misner and K.S. Thorne, Gravitation, W.H. Freeman \& Co.,
                    Princeton Uni. Press, Princeton, 1973 
\bibitem{36} M.P. Hobson, G. Efstathiou and A.N. Lasenby, General Relativity, 
                   An Introduction for Physicists, Cambridge University Press, Cambridge, 2006 
\bibitem{37} R. D'Inverno, Introducing Einstein's Relativity, Oxford University 
                   Press, Oxford and New York, 2007 
\bibitem{38} A. Banyaga, The Structure of Classical Diffeomorphism Groups, 
                    Math. \& Its Applic., Vol. 400, Kluwer Acad., Dordrecht/Boston/London, 1997
\bibitem{39} P.W. Michor and D. Mumford, Ann. Global Anal. Geom. 44 (2013) 529
\bibitem{40} J.D. Norton, Rep. Prog. Phys. 56 (1993) 791
\bibitem{41} R. Torretti, Relativity and Geometry, Dover Publications, Inc., 
                    New York, 1996
\bibitem{42} D.R. Grigore, Progr. in Phys. 47 (1999) 913 
\bibitem{43} D. Krupka, O. Krupkova and D. Saunders, Int. J. Geom. Meth. Mod. Phys. 7 
                   (2010) 631
\bibitem{44} Z.E. Musielak and R. Das, Mathematics, 13 (2025) 3928
\bibitem{45} S.W. Hawking and G.F.R. Ellis, The Large Scale Structure of Space-Time, 
                    Cambridge Uni. Press, Cambridge, 1999
\bibitem{46} V.A. Ruban, Gen, Relativ. Gravit. 33 (2001) 375
\bibitem{47} P. Gusin, A. Radosz, A.T. Augousti, J. Polonyi, O.B. Zaslavskii and R.J. Sciborski,
                    Universe, 9 (2023) 299
\bibitem{48} E. Elizalde, Phys. Rev. D 36 (1987) 1269
\bibitem{49} E. Elizalde, Phys. Rev. D 37 (1988) 2127
\bibitem{50} U.A. Yanik and K. Narayan, Class. Quantum Gravity 15 (1998) 1315
\bibitem{51} G. Tsoupros, Gen. Realativ. Gravit. 44 (2012) 309
\bibitem{52} C. Schwartz, Phys. Rev. D 25 (1982) 356
\bibitem{53} C. Schwartz, J. Math. Phys. 52 (2011) 052501
\bibitem{54} C. Schwartz, Int. J. Mod. Phys. A 31 (2016) 1650041
\bibitem{55} A. Sen, J. High Energy Phys. 204 (2002) 048
\bibitem{56} G.W. Gibbons, Phys. Lett. B 537 (2002) 1
\bibitem{57} J.S. Bagla, H.K. Jassal and T. Padmanabhan, 
                    Phys. Rev. D 47 (2003) 063504
\bibitem{58} J. Kaur, S.D. Pathak, M. Khlopov, M. Krasnov and M. Sharma, 
                    Nuclear Phys. B 1018 (2025) 117010
\bibitem{59} G.M. Murphy, Ordinary Differential Equations and Their Solutions, 
                    Dover Publ., Inc., Mineola, NY, 2011
\bibitem{60} A.M. Mathai and H.J. Haubold, Special Functions for Applied Scientists, 
                   Springer, New York, NY, 2008
\bibitem{61} A. Zee, Einstein Gravity in a Nutshell, Princeton Uni. Press, Princeton 
                   and Oxford, 2013 
\end{thebibliography}
\end{document}